# Transition from Antiferromagnetic to Quadrupole Order in a Modified Square Artificial Spin Ice

Ali Frotanpour[1,2] and Lance E. DeLong[2]

[1]Lam Research Corporation, Tualatin, Oregon, 97062

[2]Department of Physics and Astronomy, University of Kentucky, Lexington, Kentucky 40506, USA

**Abstract:**

The results of a numerical study of an ordered substitution of a double-segment into the unit cell of square artificial spin ice are presented. Each pattern vertex has three magnetic moment configurations that compete to form a magnetic ground state in zero applied magnetic field, including nearest-neighbor and next-nearest-neighbor interaction contributions to the total magnetostatic energy. The ground state depends on the number of lattice sites considered and the spacing between the two film segments that comprise the double segment. Monte Carlo simulations reveal that the double-segment sublattice adopts long-range antiferromagnetic order that can be supplanted by magnetic quadrupole order by applying an in-plane magnetic field perpendicular to the double-segment easy axis. The quadrupole-ordered state consists of a sublattice of weakly correlated chains of double-segments with antiparallel polarizations, which admits many different ground states that reflect a very high degree of frustration.

## I. INTRODUCTION

Artificial magnetic metamaterials can be designed and fabricated to display novel ordered states and phase transitions on a mesoscopic scale [1, 2]. Disconnected, 2D artificial spin ices (ASI) are dominated by dipolar interactions between semi-classical, Ising-like film segments stabilized by strong shape anisotropy. ASI were originally fabricated to imitate the frustrated behavior of 3D atomic spin ices, such as pyrochlores [3], and quadrupole ordering in $PrPb_3$ or $Ba(TiO)Cu_4(PO_4)_4$ [4-5]. From a technological point of view, ASI have significant potential for implementing tunable magnonic band structures [6] and vertex charge manipulation in reprogrammable systems [7].

Square and Kagome ASI were first introduced by Wang, et al. [2] and Qi, et al. [8] to study the validity of spin ice rules similar to those observed in 3D pyrochlore compounds. Much of the research on ASI has addressed the effects of different types of geometry and disorder on magnetic frustration and ordering, including correlated dislocations in square ASI [9], non-periodicity in quasicrystal lattices [10-14], fluctuations in thermally active systems [15, 16] and formation of superdomains [16, 17].

Recently, quadrupole order has been reported [18] in an ASI in which alternating vertical (horizontal) parallel segment pairs meet the heads of the horizontal (vertical) parallel segment pairs. Consequently, the uniformity of the magnetization of Ising segments is depressed, and the applicability of the ASI vertex concept (i.e., topological magnetic charges localized at pattern vertices) are only approximately retained.

Herein, we report a study of a square ASI modified by an ordered substitution of one of the four single-segments by a double segment, which creates a sublattice of chains of double-segments. Using the Object-Oriented Micromagnetic Framework (OOMMF), we show the modified square ASI can acquire tunable magnetic ground states. We use Monte Carlo simulations to find a field-temperature phase diagram and show that the modified square ASI undergoes antiferromagnetic order at low applied magnetic fields, and quadrupole ordering within the double-segment sublattice at higher applied fields. We find that the antiferromagnetic ordering of double-segment chains depends on the degree of order in the neighboring chains. On the other hand, quadrupole ordering involves the formation of ordered chains of double segments, where the ordering of each chain is uncorrelated with that of other chains. Consequently, the magnetic configuration of each chain can be locally manipulated without affecting other chains.

## II. MODIFIED VERTEX MODEL

The prototypical square ASI consists of fourfold vertices of disconnected Ising segments. We replaced one of the four segments with a pair of parallel segments to create an asymmetric vertex cluster shown in **Figure 1a**. The loss of fourfold symmetry can be indexed by a two-dimensional vector $\psi$, as shown in **Figure 1b**. OOMMF simulations reveal three groups of lowest-energy vertex configurations among the $2^5 = 32$ possible arrangements of the Ising states of a five-segment vertex. Ten configurations are placed among three low-energy groups shown in **Figure 1c**, and denoted Quadrupole I (TQ-I), Quadrupole II (TQ-II) and Dipole I (TD-I). (The remaining 22 configurations, assigned similar labels, are shown in the **Supplemental Information (SI), Part A.**

The energy of each vertex configuration was extracted from OOMMF simulations for different values of the spacing $d$ between the narrow Ising moments forming the double-segments. The following parameters were used in the simulation: segment length $L = 400$ *nm*, single-segment width $W_1 = 100$ *nm*, single-segment width in double-segments $W_2 = 50$ *nm*, and segment thickness $t = 10$ *nm*. Also, the permalloy saturation magnetization $M_S = 8.6 \times 10^5$ *A/m* and exchange stiffness $A = 1.3 \times 10^{11}$ *J/m³*. A suitable compromise between spatial resolution and run time is obtained by using an in-plane mesh size (as defined in OOMMF) with $\Delta x = 1$ *nm*, $\Delta y = 1$ *nm*, and a normal (to the film plane) mesh size $\Delta z = 5$ *nm*.

The OOMMF output for the vertex energies for the TQ-I, TQ-II, and TD-I configurations as a function of $d$ are shown in **Figure 2**. The configurations not shown have higher energies and can be set aside in the analyses that immediately follow (see Supplementary Information, Part A). These three configurations are very close in energy, but TD-I has the lowest vertex energy for $d > 24$ *nm*. However, the magnetostatic interaction between the two members of a double segment has lower energy in either quadruple state ($\psi = [1, -1]$ or $\psi = [-1, 1]$). Clearly, as $d$ increases, the dipolar interaction within the double segment is reduced. Note the cross-sectional area of one member of the double segment is half that of a single segment. Therefore, we have zero net magnetic charge in the TD-I configuration (unlike TQ-I or TQ-II); and consequently, TD-I has lower energy at $d > 24$ $nm$ at zero field. (In other words, the 2-in, 2-out TD-I vertex configuration minimizes head-to-tail magnetization and vertex charge). This illustrates the influence of the interaction energy between the members of the double-segments.

## III. MINIMIZING MAGNETOSTATIC ENERGY BY TILING VERTICES

We use the OOMMF vertex energies to study a Bravais lattice with a rectangular unit cell, as shown in **Figure 3a**. Note, the lattice has finite size and includes different vertices on the pattern edges and interior vertices, as shown in the SI, Tables S2, S3 and S4. We search for the overall ground state of the modified square ASI by setting $d = 24$ $nm$ and $d = 40$ $nm$.  For these $d$-values, we expect the TD-I vertex configuration to have the lowest energy, while TQ-I and TQ-II will have slightly higher energies.  We also expect the effects of next-nearest-neighbor (NNN) interactions and the system boundary to be important. The TD-I vertex is expected to generate antiferromagnetic order in the sublattice of double-segments colored red in **Figure 3b**. To confirm that TD-I vertices generate a robust ground state, we determined the lattice total magnetostatic energy as a function of the number of vertices in the horizontal direction, $n_x$.  We varied the energies for all interior vertices in the lattice by assigning them either the TD-I, TQ-I or TQ-II and configuration. Note that the configurations on the vertices on the edge will be set based on edge vertices energy levels, while the configuration of edge vertices depends on the interior vertex configurations. Each vertex energy is extracted from OOMMF as $E_d/2 + E_{ex}$ ($E_d$ is demagnetization energy and $E_{ex}$ is exchange energy), which avoids double counting of the dipolar interaction. The NNN interactions are also considered in total lattice energy calculations (details are given in the **SI, Part B**).

Since increasing the lattice size causes a huge increase in the total magnetostatic energy, it is preferable to plot the ratio of lattice energy of the TQ-I (or TQ-II) configuration to that of the TD-I configuration (e.g., $E_r \equiv E_{TQ-I}/E_{TD-I}$ ) as a function of $n_x$.

These results are shown in **Figures 4 (a)-(c)**, which also illustrates the effect of NNN interactions and vertices on the pattern edges. **Figure 4 (a)** shows that the TD-I configuration forms a ground state for $d = 24$ *nm* and $n_y = n_x$, for large enough values of $n_x$ (i.e., for an energy ratio larger than one, marked by the horizontal dashed lines in **Figure 4**). **Figure 4 (b)** gives results for $d = 40$ *nm* and $n_y = n_x$, and shows that $E_r$ increases compared to $d = 24$ *nm*, which means that the TD-I configuration becomes a stronger ground state at larger values of $d$. **Figure 4(c)** shows the effect of lattice shape by setting $n_y = 2 \times n_x$ and $d = 40$ *nm*, which creates a rectangular lattice shown in the inset to **Figure 4 (c)**. **Figure 4** indicates that $E_r$ is generally above 1.0, which indicates the TD-I configuration is a robust ground state at large enough lattice size (e.g., $n_x > 20$ and $n_y > 20$).

**Figures 4 (a)-(c)** also show the effects of ignoring interactions of vertices on the pattern edges or ignoring interactions with NNN. Removal of NNN interactions decreases $E_r$; therefore, NNN interactions favor the TD-I configuration. On the other hand, removing the energies of boundary vertices in the lattice energy calculation increases $E_r$, which indicates the interactions with the boundary tend to favor the TQ-I and TQ-II ground state configurations. Note that **Figure 3b** shows the vertices on the pattern edge (on the top and bottom) are in excited states (see **SI, Part C**).

We have also explored application of a magnetic field in the +$\boldsymbol{x}$ direction, as shown in **Figure 5a**. A Zeeman energy term modifies the lattice magnetostatic energy as follows:

$$E_{Zeeman} = -\,\Delta N_x \mu H_x \qquad (1)$$

where $\Delta N_x$ is the difference between the number of Ising segments aligned along the +$x$ and –$x$ directions, $\mu$ is the single-segment dipole moment, and $H_x$ is the applied field in the +x direction. We identified four low-energy vertex configurations in which the horizontal segments are oriented head-to-tail along the +$\mathbf{x}$ direction to minimize Zeeman energy. These are labeled as TQ-II, TQ-III, TD-II and TD-III in **Figure 5(a)**.

OOMMF simulations showed the TQ-II vertex configuration has the lowest energy, while TD-II and TQ-III have higher energies and are near-degenerate, and TD-III has the highest energy (see **SI, Part A**).  Consequently, the application of a magnetic field in +x direction favors the TQ-II configuration as a lowest energy state. Moreover, if a vertex has the TQ-II configuration, the other vertex that shares this double segment (see the yellow ellipse in **Figure 5b**) must be in the TQ-III configuration.  Therefore, it is favorable to form a TQ-II/TQ-III pair in two vertical vertex neighbors with a double segment in common. The resulting lattice configuration is shown in **Figure 5b**, where blue/purple dots highlight a pair of TQ-II/TQ-III vertices with a double segment in common.

If, on the other hand, we set all vertices into the TD-II vertex configuration, it generates a much higher total energy, since the TD-II vertex energy is higher than that of the TQ-II vertex, and is nearly degenerate with the TQ-III vertex.  As a result, the horizontal chain of double-segments exhibits quadrupole order (e.g., all double-segments in the horizontal chains are set in the $\psi = [1, -1]$ state, as shown in **Figure 5b**). Moreover, we can interchange the TQ-II vertex configuration with the TQ-III vertex configuration in the vertex pairs on a chain (see green ellipse in **Figure 5c**). This interchange does not change the total energy, and indicates there are many different lattice ground state configurations that result in a high degree degeneracy and frustration. In other words, if one chain of double segments relaxes into the $\psi = [1, -1]$ state, the other chains of double segments can remain in the $\psi = [-1, 1]$ state; therefore, the chains of quadrupoles become uncorrelated and frustrated.  Note that for a lattice size $n_y$, we can have $2^{ny}$ degenerate configurations.

## IV.  PHASE DIAGRAM

To find the phase diagram of ordered states in temperature and field, we carried out a Monte-Carlo simulation on a lattice with parameters $d$ and ratio $n_x/n_y$  (details of the Monte Carlo simulations are given in the **SI, Part D**). First, the lattice was initialized in the TD-I vertex configuration (**Figure 3b**) under zero applied field at zero temperature. Then the temperature was raised in small steps. We defined antiferromagnetic chain (AFC) order parameter was defined as:

$$AFC = \frac{1}{Nm} \sum_j | \sum_i \frac{1}{2}(-1)^i \left[ \psi_{i,j}(0) + \psi_{i,j}(1) \right] |$$ (2)

where $i$ and $j$ index the lattice sites in horizontal and vertical directions, respectively. $N$ is the total number of double-segment pairs, and $m$ is the number of lattice moment configurations (Monte Carlo trials) sampled at each temperature. The value of $\psi_{i,j}(0) + \psi_{i,j}(1)$ in Equation 2 is $\pm 2$ for the double segment dipole state, and zero for the quadrupole state. Also, the absolute value of $\psi_{i,j}(0) + \psi_{i,j}(1)$ is needed to calculate the average of the order parameter of all chains. The maximum value for the order parameter is 1.0 for a fully ordered state, and the minimum value is zero for a fully disordered state. We repeated the simulation for different values of the applied magnetic field in the **+x** direction, and monitored the value of the AFC order parameter while raising the temperature from zero.

An Ising saturated regime with the applied magnetic field in **+x** direction can be initialized with TQ-II/TQ-III pair configurations (**Figure 5b**). We executed a similar Monte Carlo simulation in which we defined the quadrupole chain order parameter (QC) as:

$$QC = \frac{1}{Nm} \sum_j | \sum_i \frac{1}{2} \left[ \psi_{i,j}(0) - \psi_{i,j}(1) \right] |$$ (3)

The value of $\psi_{i,j}(0) - \psi_{i,j}(1)$ in Equation 2 is $\pm 2$ for the double-segment quadrupole state, and zero for the dipole state. Also, the absolute value of $\psi_{i,j}(0) - \psi_{i,j}(1)$ is needed to calculate the average of chain order parameters. We repeated the simulation for different values of the applied magnetic field and monitored the QC order parameter while raising the temperature from zero.

**Figures 6 (a)-(c)** show field-temperature phase diagram for AFC ordering (left panels) and QC ordering (right panels) for the following parameters: $d = 24$ nm, $n_x = 30$, $n_y = 30$ (see **Figure 6 (a)**); for $d = 40$ nm, $n_x = 30$, $n_y = 30$ (see **Figure 6 (b)**); and for $d = 40$ nm, $n_x = 21$, $n_y = 42$ (see **Figure 6 (c)**). We adopted an order-disorder transition criterion of 0.7 [18], which revealed AFC ordering is robust at low temperatures and low fields. On the other hand, QC ordering is not apparent at low fields, but emerges at higher fields and low temperatures. The AFC ordering is gradually weakened with increasing temperature, while the QC ordering has a sharp cutoff around a specific temperature. In an intermediate region of field, the AFC and QC ordering overlap, which suggests a regime of coexisting phases near a first-order phase transition. Changes in the phase diagram for different values of $d$ or $n_y/n_x$ ratio are consistent with the growth of $E_r$ as $d$ or the $n_y/n_x$ ratio increases, as shown in **Figure 4**. Increasing $d$ from 24 $nm$ (**Figure 6(a)**) to 40 $nm$ (**Figure 6(b)**) results in a slightly larger area of AFC ordering, while the QC phase diagram area does not change, and the order parameter becomes less stable. Note the QC order parameter for $d = 40$ $nm$ gradually decreases near the cutoff of 0.7, compared to a sharp drop for QC order parameter for $d = 24$ $nm$. (indicating a possible first-order transition) Moreover, we show phase diagrams for $n_x = 21$ and $n_y = 42$ in **Figure 6(c)**, for which the total number of segments remains almost the same as for the case of $n_x = 30$ and $n_y = 30$ and the ratio of $n_y/n_x$ doubles.

It is remarkable that such a complicated phase diagram is generated by a rather simple modification of the square ASI. These results clearly suggest that we use the double-segment separation $d$ as a tuning parameter for adjusting the relative strengths of antiferromagnetic and quadrupolar correlations. For example, the quadrupole ordering can be stabilized at zero field when interior vertices are set in the TQ-I configuration and choosing smaller values of $d$, which allows us to obtain a QC ordering at relatively low field values. In this scheme, high-field, low-energy vertex configurations (i.e., a TQ-II/TQ-III configuration) suggest there is a transition from correlated double-segment chains to uncorrelated double-segment chains (see **Figure 5 (b)** and **(c)**). The reason is that the single vertical segment is frustrated; i.e., it can be magnetized in an up or down direction without changing the energy of TQ-II or TQ-III vertices. Moreover, uncorrelated chains make this ASI structure suitable for local manipulation of vertices or specific chains. Also, a careful adjustment of $d$ (~20 nm) can make the three low-energy vertex configurations nearly degenerate (see **Figure 2**), which implies the vertex configuration has 10-fold degeneracy. This situation could be exploited for locally rewritable vertices.

## V. Conclusion

In summary, we modified a square artificial spin ice to include one double segment. We used micromagnetic simulations to demonstrate a novel competition between new vertex configurations generated by the modification. We showed how the spacing between the segments in the double-segment can change the vertex ground state. We used this vertex to build a lattice structure in which a pair of vertices are correlated via a connecting double segment. We explored the effects of lattice size and segment spacing on the lattice ground state. We also considered the effect of NNN interactions in energy calculations. Moreover, we found a change in the ground state under the application of magnetic fields applied perpendicular to the double-segment. We showed that the sublattice of double-segments undergoes antiferromagnetic ordering at low magnetic field, while at higher magnetic fields quadrupole ordering takes place. Furthermore, we showed that the sublattice of double-segment chains becomes uncorrelated when field is applied in +x direction), generating a high degree of frustration while the quadrupole chain ordering remains intact. This demonstrates the lattice can be controlled and manipulated with an applied magnetic field, which is promising for engineering applications. These findings were corroborated and extended by Monte-Carlo simulations for variable temperature and field. We also identified a finite-lattice-size effect on the phase diagram, and showed that changing the ratio of $n_x/n_y$ can drastically change the phase diagram.

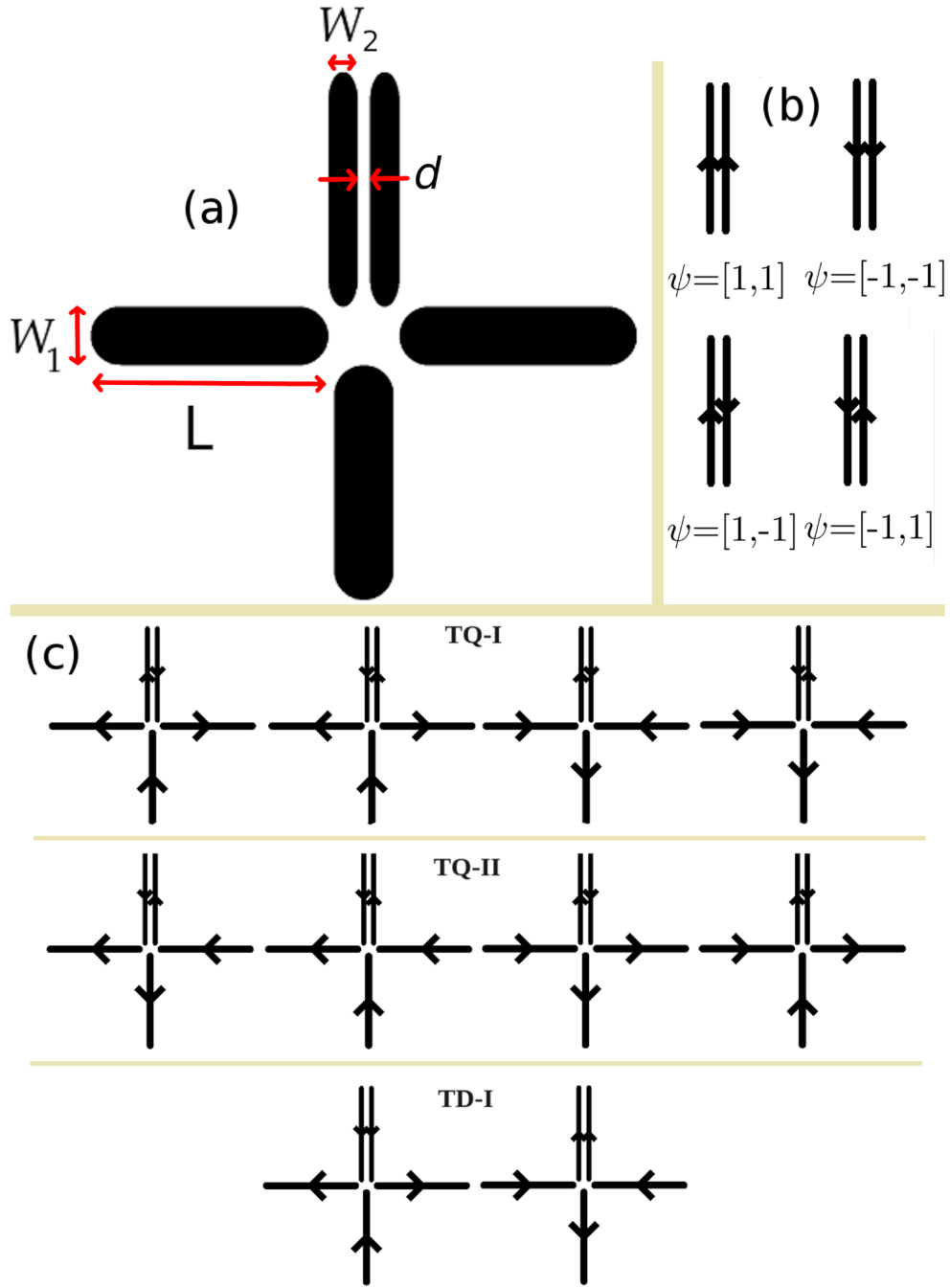

**FIG. 1.** **(a)** Modified square ASI vertex structure where *d* is the spacing between segments in the double segment. **(b)** States of the double-segment labelled by vector $\boldsymbol{\psi}$. **(c)** Three groups of low-energy moment configurations: TQ-I and TQ-II have energy degeneracy four, while TD-I has degeneracy two.

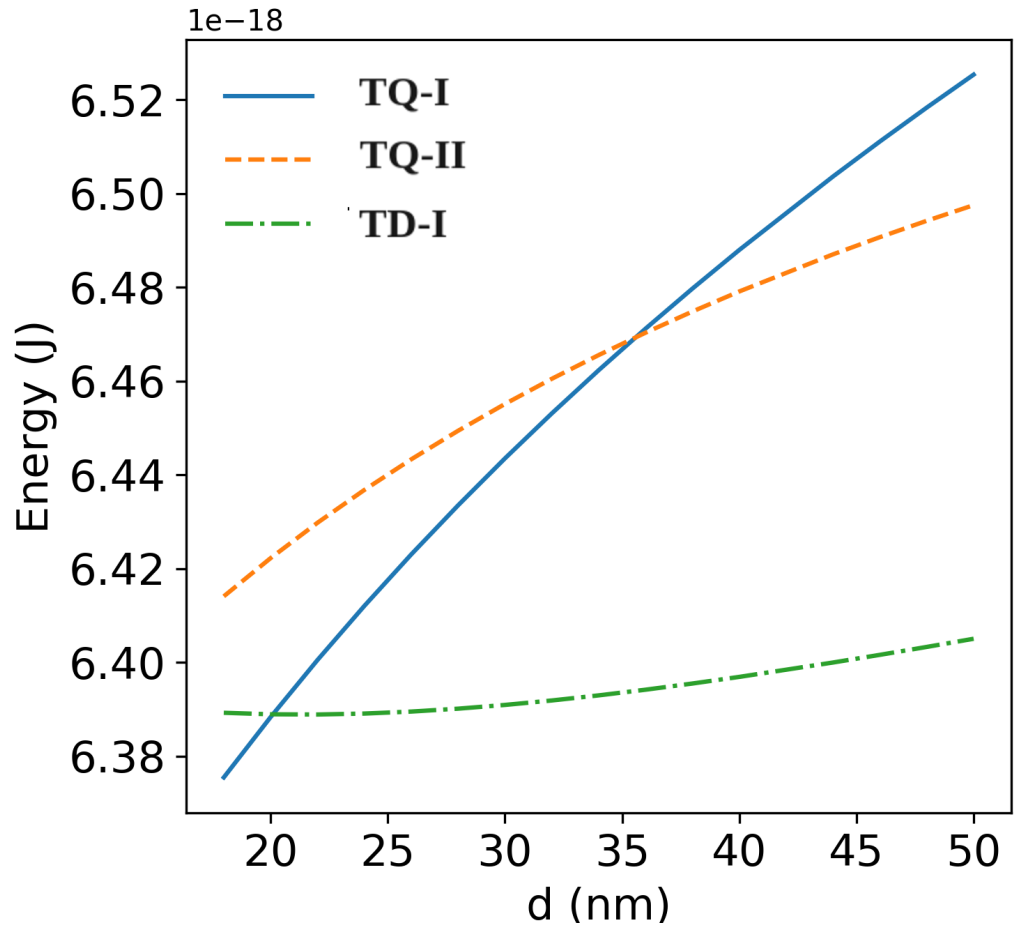

**FIG. 2.** Vertex energy as a function of the distance $d$ between segments in the double segment. Around $d = 20$ nm, TD-I and TQ-I configurations are nearly degenerate. As $d$ increases, TD-I becomes the low-energy configuration.

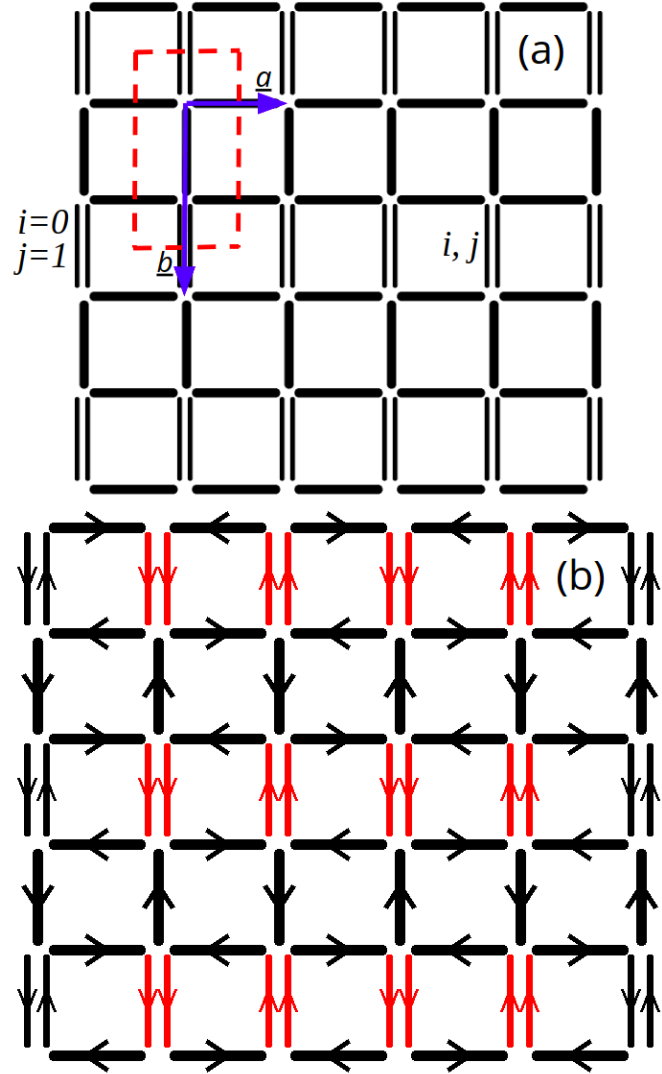

**FIG. 3. (a)** A 2D square Bravais lattice modified by adding one double-segment. The resulting rectangular unit cell is shown as the red-dashed rectangle, where **_a_** and **_b_** are primitive vectors. **(b)** Magnetic lattice formed by the TD-I vertex for a lattice size of $n_x = 6$ and $n_y = 6$. Red moments show the double-segment sublattice. We can see that each double segment forms a dipole, and the resulting sublattice undergoes long-range antiferromagnetic order. Note that each double segment on the boundary is in quadrupole state.

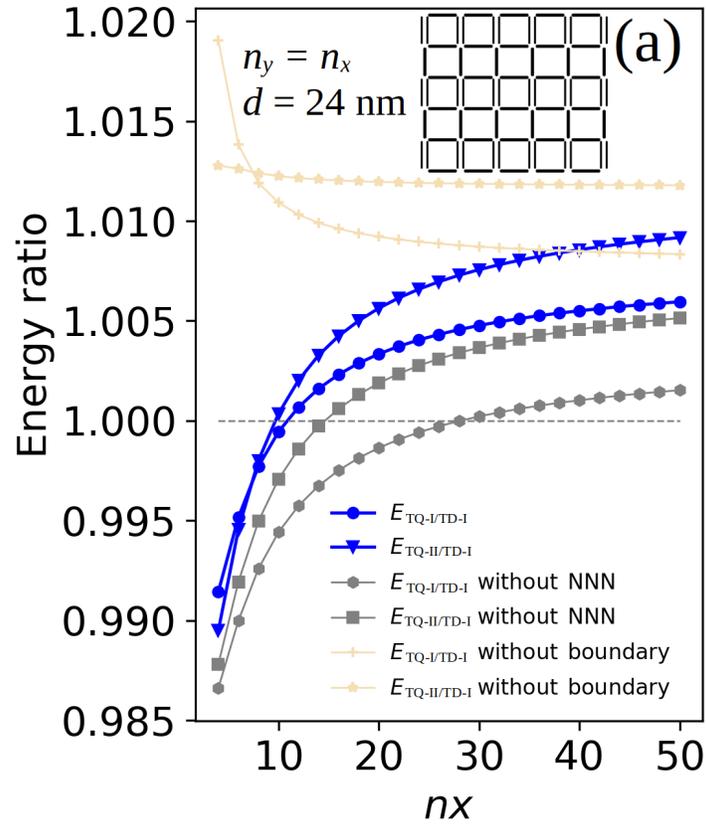

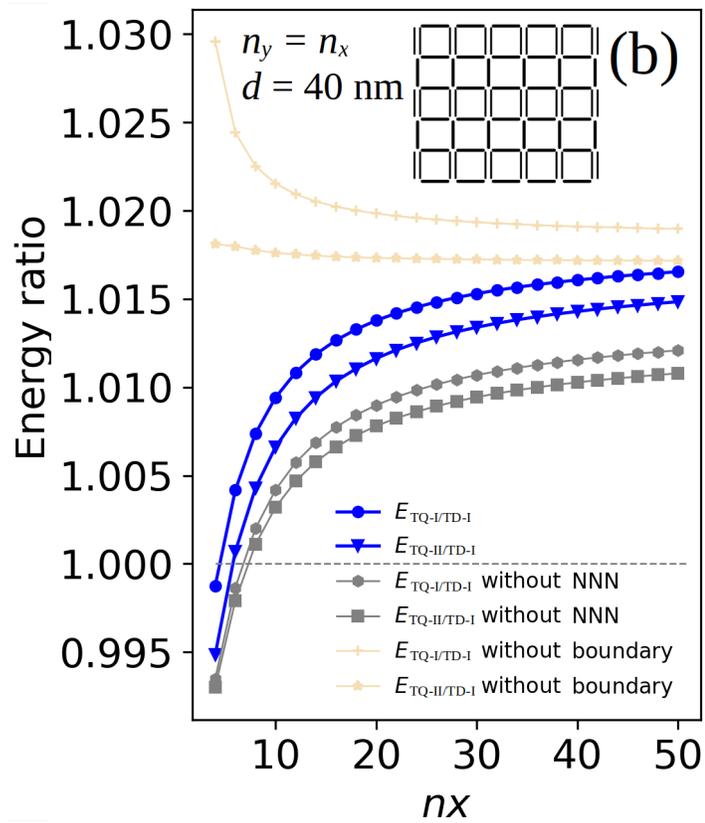

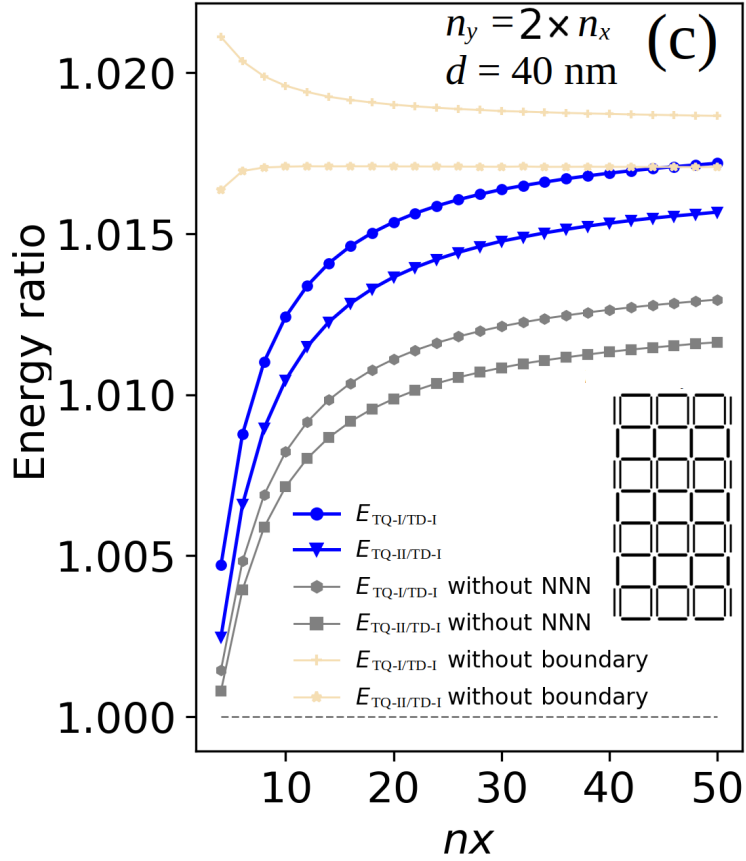

**FIG. 4.** Lattice energies of TQ-I or TQ-II vertex configurations divided by the lattice energy of the TD-I vertex configuration, as a function of lattice size $n_x$ are indicated by blue points and curves. The dashed line (energy ratio of 1) means if the curves are above it, the TD-I configuration is the ground state. Also, the effects of NNN interactions and vertices on the pattern edges can be seen by removing them from the lattice energy calculations (shown by grey and brown curves, respectively): **(a)** $n_y = n_x$ and $d = 24$ $nm$; **(b)** $n_y = n_x$ and $d = 40$ $nm$; **(c)** $n_y = 2 \times n_x$ and $d = 40$ $nm$.

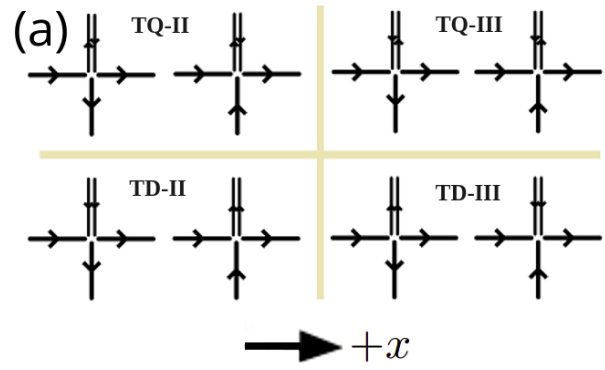

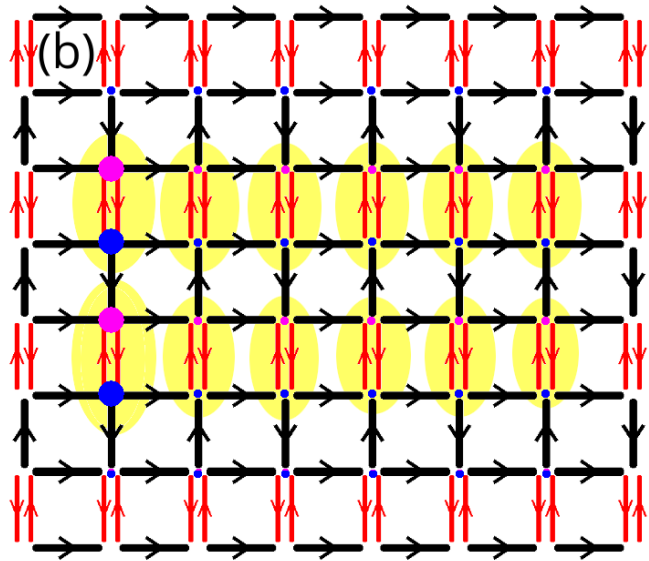

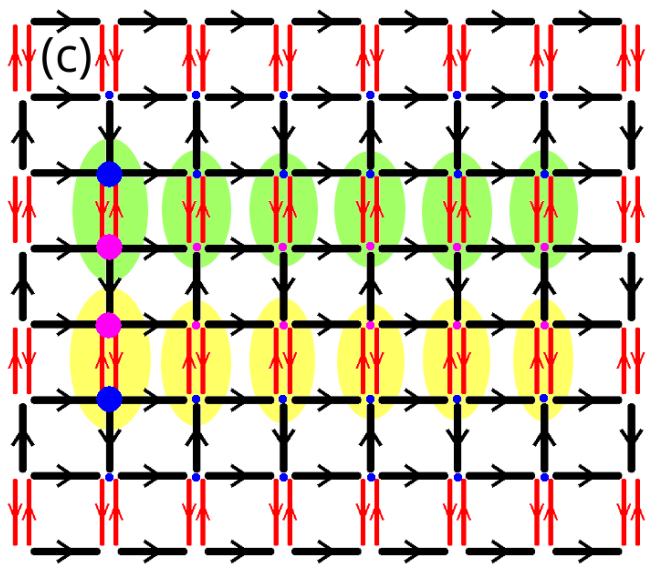

**FIG. 5.** Schematic of the Ising saturation regime. **(a)** Low-energy vertex configurations with applied magnetic field in **+x** direction (at Ising saturation). **(b)** Lattice moment configuration using TQ-II (blue dots)/TQ-III (pink dots) vertex pairs shown in the yellow region. The lattice size is $n_x = 8$ and $n_y = 8$. The double-segment sublattice is highlighted in red. **(c)** TQ-II/TQ-III (blue-pink) vertex pairs can be interchanged in each chain of double segments, without changing the energy of neighboring chains. Such a pair is shown by green ellipses.

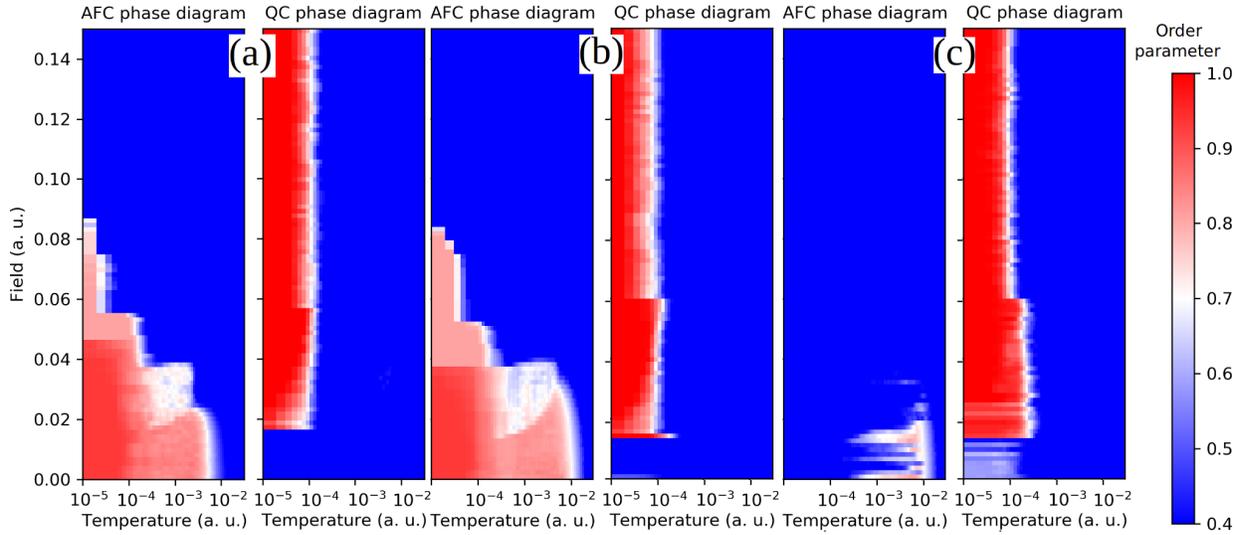

**FIG. 6.** Phase diagrams of the double-segment sublattice as a function of temperature and applied magnetic field $H_x$. Left panels show the AFC ordering diagram, and the right panels show the QC ordering diagram. The color change from red to blue indicates an order-disorder transition. The color bar on the right indicates the order parameter values. The following lattice parameters apply: **(a)** $d = 24\ nm$, $n_x = 30$, $n_y = 30$. **(b)** $d = 40\ nm$, $n_x = 30$, $n_y = 30$. **(c)** $d = 40\ nm$, $n_x = 21$, $n_y = 42$.